\documentclass[11pt]{article}
\usepackage{epsfig}

\begin{document}
\pagestyle{myheadings} \markright{On the anomalies of gravity}
\title{On the anomalies of gravity}
\author{Jos\'e B. Almeida\footnote{Universidade do Minho,
Departamento de F\'isica, 4710-057 Braga, Portugal. E-mail:
bda@fisica.uminho.pt}}

\maketitle

\begin{abstract}                
The paper is based on the recently proposed 4-dimensional optical
space theory and draws some of its consequences for gravitation.
Starting with the discussion of central movement, the paper
proceeds to establish a metric compatible with Newtonian
mechanics which can be accommodated by the new theory and finds a
correction term which can be neglected in most practical
circumstances. Being effective in the very short range, the
correction term affects substantially the results when continuous
mass distributions are considered. The main consequence is the
possibility of explaining the orbital speeds found around
galaxies, without the need to appeal for a lot of dark matter.
The speed of gravity is also discussed and the theory is found
compatible with a gravitational speed equal to the speed of
light. On the subject of black holes, it is suggested that they
are just a possibility but not a geometric inevitability.

\paragraph*{\hrulefill}
\begin{description}
  \item[Key words:]Dark matter; speed of gravity; alternative
  theories

\end{description}

\end{abstract}

\section{Introduction}
In a recent paper \cite{Almeida01} we introduced the concept of
4-dimensional optical space as an alternative to the ordinary
relativistic space. The main characteristics of the 4-dimensional
optical space being a signature 4, with the consequent Euclidean
tangential space, and geodesic arc length given by\footnote{Greek
indices take values $0$ to $3$, while latin indices take values
$1$ to $3$.}
\begin{equation}
    \label{eq:interval}
    c^2 (\mathrm{d}t)^2 = g_{\alpha \beta} \mathrm{d} x^\alpha
    \mathrm{d} x^\beta,
\end{equation}
where $c$ is the speed of light in vacuum, $t$ is time, $x^0=\tau$
is a coordinate which was named \emph{proper time} and the $x^a$
correspond to the spatial cartesian coordinates $x, y,
z$\footnote{It is customary in relativity to normalize $c=1$; in
this work we use adimensional units obtained by dividing length,
time and mass by the factors $\sqrt{G \hbar /c^3}$, $\sqrt{G
\hbar/c^5}$ and $\sqrt{\hbar c/G}$, respectively. $G$ is the
gravitational constant and $\hbar$ is Planck's constant divided by
$2 \pi$.}.

It was also stated as a premise that all particles and photons
should move along metric geodesics if a metric was chosen that
accounted for all the relevant interactions. This principle was
shown to hold for two of the main interactions in nature, namely
gravity and electromagnetism.

Although the 4-D optical space theory is expected to be
equivalent to general relativity in the prediction of test
particle trajectories and whenever trajectories follow metric
geodesics, this is not the case for massive bodies or for other
situations where relativistic trajectories deviate from geodesics.
In the present work we extract some important consequences of the
new formalism for gravitation, which explain many of the reported
puzzles of existing gravitational theory \cite{Flandern98}. The
paper is virtually self-contained, apart from concepts and
formulations that can be found in most standard textbooks; among
others we used Refs.\ \cite{Inverno96, Martin88, Schutz90}.
\section{General central movement problem}
The problem we want to solve is that of bodies orbiting other
bodies under the influence of their mutual gravitational field
but it is useful to start discussing central movement in the
context of the new theory in its most general formulation.

A general lagrangean for central movement must reflect the
combined effect of gravitational and electrostatic fields because
these are the basic interactions whose range is bigger than
atomic dimensions; however it does not need to be the most
general spherical solution. In view of the arguments presented in
\cite{Almeida01} we propose that the general form is
\begin{equation}
    \label{eq:central1}
    2L = n_\tau^2 \left(\dot{x}^0\right)^2 + n_r^2 \delta_{a b}
    \dot{x}^a \dot{x}^b = n_\tau^2 \dot{\tau}^2 + n_r^2 \left[\dot{r}^2 +
    r^2\left(\dot{\theta}^2 + \sin^2 \theta
    \dot{\phi}^2\right)\right],
\end{equation}
with
\begin{equation}
    \label{eq:spherical}
    x^0=\tau,~~ x^1 = r \sin \theta \cos \phi,~~ x^2 = r \sin \theta \sin
    \phi,~~ x^3 = r \cos \theta.
\end{equation}

The first Euler-Lagrange equation is simply,
\begin{equation}
    \label{eq:central2}
    \frac{\partial L}{\partial \dot{\tau}}
    =n_\tau^2 \dot{\tau}= \mathrm{constant}.
\end{equation}
Without loss of generality we can set $\theta = \pi/2$ because we
know beforehand that the trajectory is flat and we are at liberty
to choose the $x^3$ axis perpendicular to the plane of the
trajectory. A conservation law can be obtained from the $\phi$
equation
\begin{equation}
    \label{eq:central3}
    r n_r \dot{\phi} = J,
\end{equation}
where $J$ is a constant related to angular momentum.

Finally we write the radial equation
\begin{equation}
    \label{eq:central4}
    n_r^2 \ddot{r} = n_\tau \partial_r n_\tau \dot{\tau}^2
    + \left(r^2 n_r \partial_r n_r + r n_r^2\right)\dot{\phi}^2.
\end{equation}

For simplicity we will limit the discussion to circular orbits,
with $\dot{r}=\dot{\phi}=0$. This will quickly lead us to the
general law of angular velocity dependence on distance, without
the complications of non-circular orbits. Let us also replace the
$t$ derivative by a derivative with respect to $\tau$ by the
following relation
\begin{equation}
    \label{eq:angveloc}
    \dot{\phi} = \frac{\mathrm{d}\phi}{\mathrm{d}\tau}~\dot{\tau} =
    \omega \dot{\tau},
\end{equation}
and obtain the equation for the angular velocity in circular
motion as
\begin{equation}
    \label{eq:angveloc2}
    \omega^2 = \frac{-n_\tau \partial_r
    n_\tau}{r \left(n_r^2 + r n_r \partial_r n_r \right)}.
\end{equation}
Notice that angular velocity has been defined with respect to
proper time which coincides with time measured on the
gravitational center clock.
\section{The gravitational field}
Newtonian mechanics tells us that the gravitational pull force of
a large body, considered fixed with mass $M$, over a much smaller
body of mass $m$ is
\begin{equation}
    \label{eq:Newtonforce}
    \vec{f} = m \nabla V,
\end{equation}
where the arrows were used to denote vectors in 3D space and the
''nabla'' operator has its usual 3-dimensional meaning;  $V$ is
the gravitational potential given by
\begin{equation}
    \label{eq:Newtonpotential}
    V = \frac{G M}{r};
\end{equation}
$r$ is distance between the two bodies.

If the moving body is under the single influence of the
gravitational field, the rate of change of its momentum will equal
this force and so we write $\mathrm{d}^2\vec{r}/\mathrm{d} t^2 =
\nabla V$; $\vec{r}$ is the position vector. If we use  mass
scaling of the coordinates introduced in \cite{Almeida01} the
spatial components of the momentum must appear as $m^2 \delta_{a
b} \ddot{x}^b$; the gradient is also affected by the scaling and
we expect it to appear as $\partial_a V /m$. Using primed indices
to denote unscaled coordinates it is
\begin{equation}
    \delta_{a' b'}\ddot{x}^{b'} = \partial_{a'} V,
\end{equation}
\begin{equation}
    \label{eq:elmomentum}
    m^2 \delta_{a b}\ddot{x}^b =  \partial_a V.
\end{equation}

We are looking for a lagrangean of the type given by Eq.\
(\ref{eq:central1}), but in the absence of an electrostatic field
$n_\tau = n_r$ \cite{Almeida01}. In Cartesian coordinates it is
\begin{equation}
    \label{eq:elproposed}
    2L = n_r^2 \delta_{\alpha \beta}\dot{x}^\alpha \dot{x}^\beta.
\end{equation}
This lagrangean must be consistent with the non-relativistic form
of the gravitational force and the resulting metric must be
asymptotically flat.

If we derive the Euler-Lagrange's equations for the 3 spatial
dimensions we get
\begin{equation}
    \label{eq:eleuler}
    n_r \delta_{a \beta}\ddot{x}^\beta = \partial_a
    n_r \delta_{\alpha \beta}\dot{x}^\alpha\dot{x}^\beta
    = \partial_a n_r \left[\left(\dot{x}^0\right)^2
    + \delta_{i j} \dot{x}^i \dot{x}^j\right].
\end{equation}
From Eq.\ (\ref{eq:elproposed}) we take $ \delta_{i j}\dot{x}^i
\dot{x}^j=1/n_r^2 - (\dot{x}^0)^2$, to be replaced above:
\begin{equation}
    \label{eq:eleuler2}
    n_r^2 \delta_{a \beta}\ddot{x}^\beta = \frac{\partial_a n_r}{n_r}.
\end{equation}

It is now convenient to make the replacement $n_r = m \eta_r$, so
that the previous equation becomes
\begin{equation}
    \label{eq:eleuler3}
    m^2 \eta_r^2 \delta_{a \beta}\ddot{x}^\beta
    = \frac{\partial_a \eta_r}{\eta_r} = \dot{p}_a .
\end{equation}

If Eq.\ (\ref{eq:eleuler3}) is to produce the same results as
Eq.\ (\ref{eq:elmomentum}) at appreciable distances from the
central body, $\eta_r$ must be a function of $r$ that when
expanded in series of $1/r$ has the first two terms $1 + M/r$,
where adimensional units were used to remove the gravitational
constant. An interesting possibility is the function
\begin{equation}
    \label{eq:gravindex}
    \eta_r = \mathrm{e}^V = \mathrm{e}^{M/r}.
\end{equation}
The second members of the two equations are now equal and the
first members will be equivalent in non-relativistic situations.
So compatibility with Newtonian mechanics is ensured.

\section{Angular velocity dependence on distance}
We are now in position to discuss how the angular velocity of
orbiting bodies is expected to depend on the distance to the
central mass. Our reference is the $r^{-3/2}$ dependence from
Newtonian gravitation but we wish to find some explanation for
the apparent anomalies near galaxies \cite{Flandern98}, which
people usually attribute to large amounts of dark matter.

We go back to Eq.\ (\ref{eq:angveloc2}) and replace both $n_\tau$
 and $n_r$ by $m^2 \eta_r$ to get
\begin{equation}
    \label{eq:angrelat1}
    \omega^2 = \frac{M}{r^3 + M r}.
\end{equation}
The angular velocity is shown to depend on the distance in a
similar way to the Newtonian predictions at large distances, but
closer to the central mass the dependence evolves gradually to
$\sqrt{r}$. The range of the correcting term can be found by
setting $M/r^2 = 1$; for the Sun this range is $5 \times
10^{-16}~\mathrm{m}$. This effect becomes more important as
masses grow and a range of $1~\mathrm{m}$ would imply a central
mass of $8 \times 10^{61}~\mathrm{Kg}$, or about $4 \times
10^{31}$ times Sun's mass.

Naturally, being a short range correction, the $1/\sqrt{r}$
dependence must be decisive in the way continuous mass
distributions behave. A correct study of continuous mass
distribution dynamics is a difficult subject, which we can turn
into a more manageable one by assuming some law for the mass
density. This posture is not unquestionable because the net
effect of a mass element will be different whether it is
stationary or moving relative to the center of mass. In spite of
these limitations, we think it is very enlightening to make
estimates based on pre-defined density laws.

As usual we will assume that the mass distributions are
spherically symmetrical or symmetrical about one axis. When
studying the movement of a mass element we only have to consider
the mass inside of a sphere with radius equal to the distance of
the mass element to the center of mass, for spherically
symmetrical distributions, or the mass of a cylinder of the same
radius, for the axis symmetric distributions.

If the mass density is $\rho$, the total mass inside a sphere of
radius $r$ is given by
\begin{equation}
    \label{eq:masssphere}
    M = 4 \pi \int_0^r \rho {r'}^2 \mathrm{d}r',
\end{equation}
while for a cylinder with the same radius we have
\begin{equation}
    \label{eq:masscyl}
    M = 2 \pi \int_0^r \rho r' \mathrm{d}r'.
\end{equation}

We will now look at two special cases: The first one has a mass
density decaying with $1/r^2$ and spherical distribution while in
the second one mass density decays with $1/r$ and symmetry is
axial. The previous equations become respectively
\begin{equation}
    \label{eq:masssphere2}
    M = 4 \pi \rho_0 r,
\end{equation}
\begin{equation}
    \label{eq:masscyl2}
    M = 2 \pi \rho_0 r,
\end{equation}
with $\rho_0$ some constant. In both cases the mass affecting the
orbit of the mass element at distance $r$ can be expressed as
$M_0 r$. If this is substituted in Eq.\ (\ref{eq:gravindex}) the
resultant $\eta_r$ is a constant and the corresponding angular
velocity is zero.

Let us look at what Newtonian mechanics predicts for the same
situations: We take the expression for angular velocity $\omega^2
= M / r^3$ and replace $M$ by $M_0 r$, to get an angular velocity
that decreases linearly with $1/r$. This difference is of the
utmost importance when studying the dynamics around galaxies. If
beyond some distance the mass density of a spiral or disk shaped
galaxy shows a dependence on $1/r$, the angular velocity will be
governed only by the mass in the core of the galaxy. Given the
rough approximations made in this study it is conceivable that
some density law can even counteract this dependence yielding an
almost constant angular velocity, without the need to call for
dark matter.
\section{Speed of gravity and black holes}
In general relativity gravity must act instantaneously
\cite{Flandern98}, which is seen as a contradiction with
Einstein's philosophy. It appears as information being carried at
cosmological distances in no time and this is certainly not in
accord with the spirit of relativity.

Eq.\ (\ref{eq:angveloc2}) holds the key to the paradox. Angular
velocity is here defined relative to proper time and proper time
is synchronized all over the Universe by photons
\cite{Almeida01}; whatever carries gravity must then travel at
the speed of light. In a forthcoming paper we will show that
gravity is indeed intermediated by so called \emph{gravitons},
which are just another view of photons. We will also show
experimental proof of graviton existence and detection.

And what about black holes? Black holes can be accommodated in 4D
optical space  by the consideration of $\eta_r$ given by
\begin{equation}
    \label{eq:schwart}
    \frac{1}{\eta_r} = 1 - \frac{M}{r}.
\end{equation}
Trajectories evaluated with the metric resulting from the
definition above will almost duplicate Schwartzschild's
geodesics, except for different angular velocities. On the other
hand, there is no geometrical reason to impose this solution, as
was the case with general relativity \cite{Schwart16}, and we
think it is good to keep an open mind but until there is evidence
of their existence it is best not to consider black holes at all.
\section{Conclusion}
All over the past year the author has been addictive to the idea
of a new formulation of general relativity, motivated by
similarities with optical propagation. He has been tormenting
colleagues and friends with his ideas, which were made public at
the OSA meeting of October 2000 \cite{Almeida00:4}. The first
concepts were developed into a theory \emph{named 4-dimensional
optical space theory} and written in the form of a paper
\cite{Almeida01} in last April. The present work is the first to
extract new results from this theory in order to seek an
explanation for existing puzzles in the Universe.

It has been shown that the new theory is compatible with
established results from Newtonian gravitation and is more
effective than relativity in generalizing them to the
cosmological scale. In particular it was possible to explain
observed anomalies in the gravitation around galaxies without the
need to appeal for large amounts of dark matter. The paper was
also successful in reconciling a propagation speed for relativity
with the stable orbits of planets, without the consequence of an
inevitable fall into the Sun. As a final result a discussion of
black holes showed that although these can be accommodated by the
new theory they are not inevitable and can be left as an open
question pending observational evidence.

The author believes that the present work definitely establishes
the new theory as the best existing theory for gravitation but he
has much wider expectations. Some of his still unpublished
results show that gravity is quantized and intermediated by
massless particles, analogous to photons; other results show that
quantum mechanics follows directly from the theory. These results
will be the subject of forthcoming papers. Ultimately the author
hopes to include strong interaction and find mass as a
consequence of geometry.

%


  \bibliography{aberrations}   
  \bibliographystyle{unsrt}

\end{document}